\documentclass[aps,pre,reprint,unsortedaddress]{revtex4-2}
\usepackage{hyperref}
\usepackage{lipsum}
\usepackage[caption=false]{subfig}
\usepackage{graphicx}
\usepackage[utf8]{inputenc}
\usepackage[T1]{fontenc}
\usepackage{amsmath}
\usepackage{amssymb}
\usepackage{graphicx}
\usepackage{siunitx}
\usepackage{mathtools}
\usepackage{booktabs}
\usepackage{lineno}

\newcommand{\uin}{\text{IN}}
\newcommand{\uout}{\text{OUT}}

\hypersetup{%
    pdfpagemode={UseOutlines},
    bookmarksopen,
    pdfstartview={FitH},
    colorlinks,
    linkcolor={black},
    citecolor={black},
    urlcolor={blue}
  }

\begin{document}


\title{Wall-induced anisotropy effects on turbulent mixing in channel flow: A network-based analysis}

\author{D. Perrone}
\email[]{davide.perrone@polito.it}
\affiliation{Department of Mechanical and Aerospace Engineering, Politecnico di Torino, 10129 Turin, Italy}
%
\author{J.G.M. Kuerten}
\affiliation{Department of Mechanical Engineering, Eindhoven University of Technology,
P.O. Box 513, 5600 MB Eindhoven, The Netherlands}

\author{L. Ridolfi}
\affiliation{Department of Environmental, Land and Infrastructure Engineering, Politecnico di Torino, 10129 Turin, Italy}

\author{S. Scarsoglio}
\affiliation{Department of Mechanical and Aerospace Engineering, Politecnico di Torino, 10129 Turin, Italy}

\date{\today}

\begin{abstract}

Turbulent mixing is studied in the Lagrangian framework with an approach based on the complex network formalism. We consider the motion of passive, noninertial particles inside a turbulent channel simulated at $Re_{\tau} = 950$. The time-dependent network is built to evaluate the transfer of tracers between thin wall-parallel layers which partition the channel in the wall-normal direction. By doing so, we are able to assess the spatial and temporal complexities arising from turbulence dynamics and their influence on the mixing process. This approach highlights the effects of small-scale features of turbulent flow structures and also the larger scale effects determined by wall-induced anisotropy. Complex networks, coupled to the Lagrangian description of turbulence, are effective in providing novel insights into inhomogeneous turbulence and mixing.
\end{abstract}

\maketitle

\section{Introduction}

Mixing processes permeate several physical phenomena and engineering applications, such as the interaction of chemical species transported in a fluid or the dispersion of pollutant in the atmosphere \cite{dimotakis2005turbulent,shraiman2000scalar}. Turbulence greatly enhances mixing; when a contaminant is inserted into a high Reynolds number flow, the stirring action of eddies thoroughly mixes the fluid with the contaminant, until the average of the spatial concentration gradients has vanished. The Lagrangian formalism, which relies on following the trajectories of fluid particles, is particularly suitable to represent mixing problems. A diffusing species can be modeled by Lagrangian fluid particles and therefore treated as a tracer which is simply advected by the flow under three main conditions. The species, or \textit{scalar}, must be passive, \textit{i.e.}, its motion and concentration do not exert feedback on the flow; its response time must be much smaller than the smallest dynamically relevant timescale in the flow (its Stokes number must be much smaller than one); finally, the scalar's own molecular diffusion must be negligible compared to the turbulent diffusion, which is usually true in turbulent flows \cite{stelzenmuller2017lagrangian,polanco2019lagrangian}.

Past research on tracer dynamics was devoted to investigate single and multi-particle statistics, providing results on the asymptotic behavior of tracers \cite{taylor_diff}, their pairwise separation \cite{salazar2009two,polanco2018relative}, the shape evolution of higher order particle sets \cite{biferale2005multiparticle,bianchi2016evolution} and Lagrangian coherent structures \cite{schneide0,schneide1}. 

A thorough description of mixing requires to fully retain the spatial and temporal complexity of the turbulent flow. This work aims to introduce tools, based on complex networks, to complement and enhance the results from statistical analysis. In particular, the effects of spatial inhomogeneities and the initial transient phase of dispersion pose significant challenges to a complete description of mixing \cite{sreenivasan2019turbulent}. 

To gain a deeper insight into mixing processes in inhomogeneous turbulence, we considered the motion of numerically generated tracers in a direct numerical simulation of a channel flow at a frictional Reynolds number $Re_{\tau} = 950$. After partitioning the channel in fixed layers along the wall-normal direction, we analyzed the trajectories and kept track of the partitions they visited. A similar approach, based on the discretization of the transfer operator, has been already applied to dynamical systems and geophysical flows \cite{froyland2005statistically,dellnitz2009seasonal,sergiac}. The resulting representation of the mixing process is a complex network. The rationale for doing so is twofold: first, the network-based approach enables us to obtain a condensed yet complete representation of the mixing process, without losing spatial or temporal information; and second, we are able to use established tools from graph theory, which are particularly suitable to represent large sets of dynamical interacting units. The network approach provides us with adequate measures to identify the onset of the asymptotical dispersion regime while also retaining the complexity of the transient stage. We also exploit the transport network representation to measure the progress of the mixing process, paying particular attention to the anisotropic behavior due to wall shear. 

Complex networks have proven valid to study a diverse range of natural phenomena \cite{boccaletti}; classically, these methods have been used to describe, for example, networks of social interactions or transportation systems \cite{costa2011analyzing}. Recent approaches have focused on fluid dynamics and, in particular, turbulence. Applications have explored the Eulerian reference frame with the analysis of time series, multipoint correlations and vortex dynamics \cite{scarsoglio2016complex,iacobello2018visibility,taira16,meena18, bai19}; other studies focused on Lagrangian trajectories, building networks that take into account mutual separation of particles \cite{schlueter2017coherent,schlu17,schlueter19,iacobello2019lagrangian,padberg17,rypina17}.

The present work is divided into four sections. After this introduction, section \ref{sec:data} describes the method employed to obtain the trajectory data, with further details about the direct numerical simulations given in the Appendix. Section \ref{sec:net} introduces the concept of network and defines some metrics, with particular focus on directed and weighted networks. Section \ref{sec:def} reports the definition of the transport network which is the subject of the present work and some of its basic characteristics.
Section \ref{sec:res} contains the application of the network to the trajectories and results concerning the temporal evolution of the network (section \ref{sec:weight}), the number of connections and their distribution in time and space (section \ref{sec:metric}) and the application of an algorithm to partition the channel (section \ref{sec:com}). Finally, a discussion and concluding remarks are given in section \ref{sec:concl}.

\section{Channel simulation and network definition}
\subsection{Lagrangian data}
\label{sec:data}

In order to obtain a set of Lagrangian trajectories we exploited the direct numerical simulations (DNS) performed by \textcite{kuerten13}, which comprise the turbulent channel flow simulation and the integration scheme for trajectories. A fully developed and incompressible turbulent flow was simulated by numerically solving the Navier-Stokes equations inside a rectangular box of size $2\pi\delta\times 2\delta \times \pi\delta$, where $\delta$ is half the channel height. The frictional Reynolds number is $Re_{\tau} = \delta u_{\tau}/\nu = 950$, where $\nu$ is the kinematic viscosity and $u_{\tau} = \sqrt{\tau_w/\rho}$ is the friction velocity, with $\tau_w$ being the shear stress at the walls and $\rho$ is the mass density of the fluid. $Re_{\tau}$ is kept fixed by prescribing the mean driving force in the $x$ direction. 

After the flow variables achieved statistical convergence, $N_p = \num{10000}$ passive tracers were seeded inside the channel in a square grid at $x^+ = x\nu/u_{\tau} = 0$ and their trajectories were integrated through the total simulation time $T_{\text{DNS}}$, whose value in wall units is $T^+_{\text{DNS}} = T_{\text{DNS}} u_{\tau}^2/\nu = \num{15200}$ (throughout the paper the $^+$ superscript indicates wall unit normalization). Since these particles are assumed massless, their Lagrangian velocity $\mathbf{v}(\mathbf{x}_0, t)$ matches at any time $t$ the velocity of the underlying Eulerian field $\mathbf{v}(\mathbf{x}, t)$, \textit{i.e.} $\mathbf{v}(\mathbf{x}_0, t) = \mathbf{v}(\mathbf{x}(\mathbf{x}_0, t),t)$, where $\mathbf{x}_0$ is the starting coordinate of each tracer and $\mathbf{x}(\mathbf{x}_0, t)$ is the subsequent trajectory. The velocity at the location of tracers was obtained interpolating the Eulerian velocity field resulting from the simulation; the integration was performed using the same temporal scheme as employed for the solution of the Navier-Stokes equations.
The tracers release pattern is a $100\times 100$ square grid, as exemplified in the left side of figure \ref{fig1}a, equispaced in both the $y$ and $z$ directions; the spacing along $y$ is $\Delta y^+ = 19$ and the first layer is at a distance $y^+_0 = 9.5$ from the wall, while the spacing along $z$ is $\Delta z^+ = 29.85$. The wall normal release coordinate is then $y^+_0 = 9.5 + 19(i-1)$, where $i$ is the starting layer. Also, we usually show the distance from the nearest wall - ranging from $y^+ = 0$ at the walls to $y^+ = 950$ at the centerline - instead of the standard $y^+$ coordinate; by doing so, we can refer to both near-wall regions as low-$y^+$ regions.

Additional details on the direct numerical simulation and the interpolation scheme are given in the Appendix.

\subsection{Networks: definition and metrics}
\label{sec:net}

In this section, a brief introduction on networks and their metrics is given, with a focus on directed and weighted graphs; these tools will be used in the following to describe mixing in channel flow. A network $G(\mathcal{V}, \mathcal{E})$ contains a set $\mathcal{V}$ of $N$ interacting objects (\textit{nodes}, or \textit{vertices}) and a set $ \mathcal{E}$ of interactions (\textit{links}, or \textit{edges}). In a weighted and directed graph, each link $\left\lbrace i,\,j\right\rbrace \in \mathcal{E}$ connects an ordered pair of vertices $i$ and $j$ and has an associated weight $W_{ij}$, which may represent the strength of the interaction, the physical distance between nodes or other features of the network \cite{boccaletti}. Also, one may want to consider links starting and ending in the same node, so that  $\left\lbrace i,\,i\right\rbrace \in \mathcal{E}$; these links are called \textit{loops} and the resulting graph is more properly defined as a multigraph \cite{bollobas2013modern}. The whole set of links can be represented in a compact way by means of the \textit{adjacency matrix} $\mathbf{A}$, an $N\times N$ matrix defined as
\begin{equation}
\label{eq:adj}
A_{ij} = \left\lbrace\begin{matrix}
1\text{ if } \left\lbrace i,\,j\right\rbrace \in \mathcal{E} \\
0\text{ if } \left\lbrace i,\,j\right\rbrace \notin \mathcal{E} .\\
\end{matrix} \right.
\end{equation}
We note that in a directed network $\lbrace i,\,j\rbrace \in \mathcal{E}$ does not imply $\lbrace j,\,i\rbrace \in \mathcal{E}$, since each connection contains directional information. Similarly, a weight matrix $\mathbf{W}$ can be defined as a matrix with elements $W_{ij}$ equal to the weight of the link between vertices $i$ and $j$ ($W_{ij} = 0$ if $\left\lbrace i,\,j\right\rbrace \notin \mathcal{E}$). Because of the directionality of links, neither $\mathbf{A}$ nor $\mathbf{W}$ are symmetrical for a directed graph. The number of connections incident to each node $i$ is the \textit{degree} $k_i$ of that node; for directed networks it is useful to define both an ingoing degree $k_i^{\uin}$ and an outgoing degree $k_i^{\uout}$, which are the number of links entering the $i$-th node or leaving it, respectively. Their calculation starting from $\mathbf{A}$ is performed as follows:
\begin{equation}
\label{eq:k}
k_i^{\text{IN}} = \sum_{j\in \mathcal{V}} A_{ji}\quad\quad k_i^{\text{OUT}} = \sum_{j\in \mathcal{V}} A_{ij}.
\end{equation}
For weighted networks, it may also be useful to consider the sum of incident (both ingoing and outgoing) weights; thus the strength $s_i$ of nodes can be defined:
\begin{equation}
\label{eq:s}
s_i^{\text{IN}} = \sum_{j\in \mathcal{V}} W_{ji}\quad\quad  s_i^{\text{OUT}} = \sum_{j\in \mathcal{V}} W_{ij}.
\end{equation} 

Both degree and strength are measures of node centrality, which is the importance of a vertex in a graph. A different metric for centrality is the \textit{betweenness} $B_i$, which is defined as
\begin{equation}
\label{eq:bet}
B_i = \sum_{s,t\in\mathcal{V}\,s\neq t} \frac{n_{st}(i)}{n_{st}},
\end{equation}
where $n_{st}$ is the total number of shortest paths between nodes $s$ and $t$ and $n_{st}(i)$ is the number of those paths that pass through node $i$. This metric measures the importance of a node not by the number of its connections, but rather by its influence on the flow of information between different nodes \cite{girvan2002community}. Its definition can be extended to measure the number of shortest paths running through an edge, thus defining the \textit{edge betweenness} $B_{ij}$ for each link $\lbrace i,\, j \rbrace \in \mathcal{E}$ as
\begin{equation}
\label{eq:betedge}
B_{ij} = \sum_{s,t\in\mathcal{V}\,s\neq t} \frac{n_{st}(i,j)}{n_{st}},
\end{equation}
where, similarly to equation \eqref{eq:bet}, $n_{st}(i,j)$ is the number of shortest paths from node $s$ to node $t$ that pass through link $\lbrace i,\, j \rbrace$.

When each node has a defined position in Cartesian space (as is our case), a physical length $d_{ij}$ can be associated to each link, equal to the Euclidean distance between interacting nodes. The product $d_{ij}W_{ij}$ of a link's length and its weight can be used to jointly represent the intensity and distance of a connection and, \textit{e.g.}, highlight long distance connections that carry significant information. To sum up weighted physical length information incident to a single node, we define the \textit{mean weighted link length} $D_w$ of the $i$-th node as
\begin{equation}
\label{eq:wedgel}
D_{w,i}^{\uin} = \frac{1}{k_i^{\uin}}\sum_{j\in\mathcal{V}} d_{ji}W_{ji},\quad D_{w,i}^{\uout} = \frac{1}{k_i^{\uout}}\sum_{j\in\mathcal{V}} d_{ij}W_{ij},
\end{equation}
which is the average of all weighted link lengths incident to a node $i$, taken over ingoing and outgoing links separately.

Finally, a useful tool to detect communities inside a network is the modularity $Q$, introduced by \textcite{newman2004finding}, which is a measure of the strength of a given graph partition. A community is a group of vertices which have a large number of links between themselves and only few running outside the community. The value of the modularity is the fraction of link weights that lie inside a community minus the same number if they were disposed randomly. Its definition for weighted and directed networks is \cite{newman2004weighted,leicht2008community,nicosia2009extending}
\begin{equation}
\label{eq:mod}
Q = \frac{1}{m}\sum_{i,j} \left[W_{ij} - \frac{s_i^{\uin}s_j^{\uout}}{m}\right]\delta_{c(i),c(j)},
\end{equation}
where $m = \sum_{ij} W_{ij}$ is the sum of all link weights and $\delta_{c(i),c(j)}$ is equal to one if vertices $i$ and $j$ are in the same community and zero otherwise. The contribution $q_k$ of the $k$-th community to the overall modularity can be obtained by restricting the summation in equation \eqref{eq:mod} to a single community; of course, the sum of all contributions is the modularity $Q = \sum_k q_k$.

\subsection{Transport network definition}
\label{sec:def}

\begin{figure}
\includegraphics[width = 8.6cm]{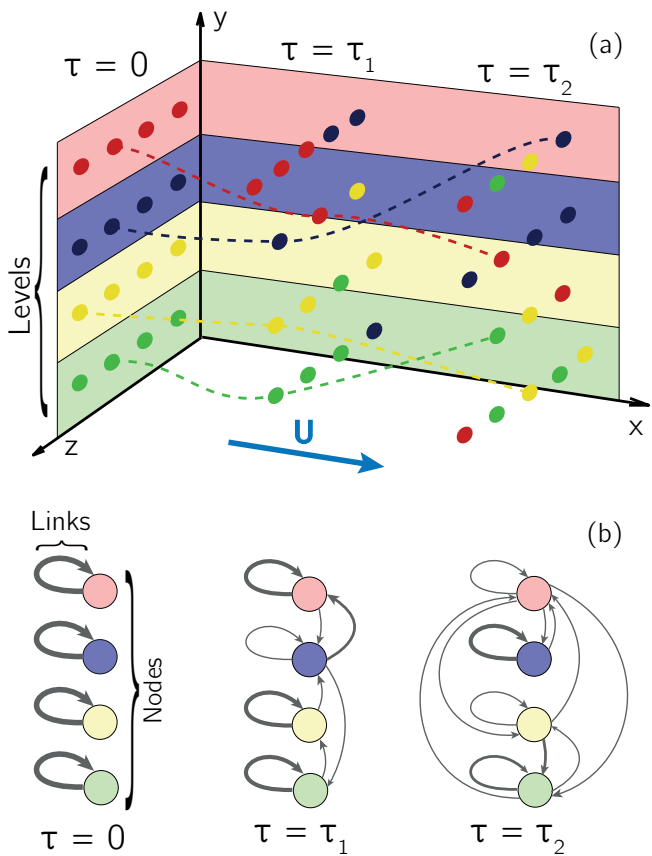}
\caption{(a) The channel is divided into levels, here identified by the colored bands, and each level is represented by a node of the network. Tracers are marked with their starting levels and are released in a square grid at $x^+ = 0$; as they move across the channel they enter different levels, thus establishing connections from their starting node to the present ones. (b) The network nodes represent the discrete levels of the channel in the same order as reported in panel (a), while links are determined by the exchange of tracers; here links are depicted with thickness proportional to the link weight, \textit{i.e.} the number of exchanged tracers. As time proceeds, connections between farther and farther levels are established. \label{fig1}}
\end{figure}

Aiming to quantify the nature of the fluid exchanges between different regions of the flow, we built a network starting from the trajectory data, with a focus on processes arising from the spatial inhomogeneity of the channel. We defined $N_l = 100$ levels that partition the channel in the inhomogeneous $y$ direction; each level has a constant height $\Delta y^+ = 19$. By doing so, we address mixing phenomena induced by inhomogeneities caused by the channel boundaries. The fluid in the channel moves between different levels, which are represented by nodes in the network; we quantify this motion by measuring the number of tracers released at $t^+ = 0$ in level $i$ that at a given time $\tau^+$ are in level $j$ (see figure \ref{fig1}a). This translates in a direct way into a time-dependent definition of link for the network: two nodes $i$ and $j$ are connected at time $\tau^+$ if a tracer, released in level $i$, is located in level $j$ at $t^+ = \tau^+$ (see figure \ref{fig1}b). As can be seen in the figure, when particles are released ($t^+ = 0$) the network is trivial, containing only self-loops; as time grows, tracers move farther and farther away from their starting level and new, longer range connections are activated. Moreover, we set the weight of this link to be proportional to the number of exchanged tracers. The resulting network is weighted and directed, since it retains a directional information on the motion of tracers. It is also spatially-embedded - as nodes have a physical location inside the channel which is identified by their $y^+$ coordinate - and time-dependent, because the existence and weight of the links depend on the delay $\tau^+$ between particle release and the time at which the network is defined. Since we do consider loops in the network definition, \textit{i.e.} links corresponding to tracers which do not change level over time, the transport network is a multigraph; the metrics defined in section \ref{sec:net} can be applied unchanged including these loops. 

The entire set of links can be formally represented as a weight matrix $\mathbf{W}$, which is built starting from the link definition as
\begin{equation}
\label{eq:W}
W_{ij}(\tau^+) = \frac{N_{i\to j}(\tau^+)}{N_i(t^+ = 0)},
\end{equation}

where $N_{i\to j}(\tau^+)$ is the number of tracers, originally located in level $i$, which are in level $j$ at time $t^+ = \tau^+$; $N_i(t^+ = 0)$ is the number of tracers present in level $i$ at the start of the simulation, which is equal to 100 for all levels. The weight matrix is time-dependent and asymmetric ($W_{ij}\neq W_{ji}$) because of the network directionality; for this reason we are able to define ingoing and outgoing metrics, accounting for the motion of tracers into and out of levels. The rows of $\mathbf{W}$ always add up to 1; from a probabilistic point of view, each of its elements $W_{ij}(\tau^+)$ can be interpreted as the probability that a tracer, starting from level $i$, is located in level $j$ at time $\tau^+$. The summation of the columns of $\mathbf{W}(\tau^+)$ is equal to the number of tracers present in each level at time $\tau^+$ divided by $N_i(t^+ = 0) = 100$. Some network metrics, in particular the degree centrality, require also the use of the adjacency matrix $\mathbf{A}$, which we define for the transport network as 
\begin{equation}
\label{eq:Atransport}
A_{ij}(\tau^+) = \left\lbrace\begin{matrix}
1\text{ if } W_{ij}(\tau^+) \neq 0 \\
0\text{ if } W_{ij}(\tau^+) = 0 .\\
\end{matrix} \right.
\end{equation}

The transport network allows us to fully describe the vertical mixing in a discrete approach. Moreover, although a single transport network $\mathbf{W}(\tau^+)$ only accounts for the start and end position of each tracer after a delay $\tau^+$, analyzing the network evolution for a wide interval of times provides a complete picture of the temporal evolution of mixing. On the other hand, diffusion in the homogeneous directions was neglected, as was motion with a typical scale smaller than the node height, namely $\Delta y^+ = 19$. As an example, the two levels nearest to the top and bottom walls comprise both the viscous and buffer sublayers, making it impossible to distinguish between the two using this approach. 

\section{Results}
\label{sec:res}
We now discuss the results obtained by applying the transport network to the set of trajectories. The following analyses were carried out by transforming the particle dynamics into its matrix representation via the complex network formalism. This step simplifies the investigation of particle motion, while retaining most of its features in the $y^+$ direction.

\subsection{Transport matrix evolution}
\label{sec:weight}
\begin{figure*}
\includegraphics[width = \textwidth]{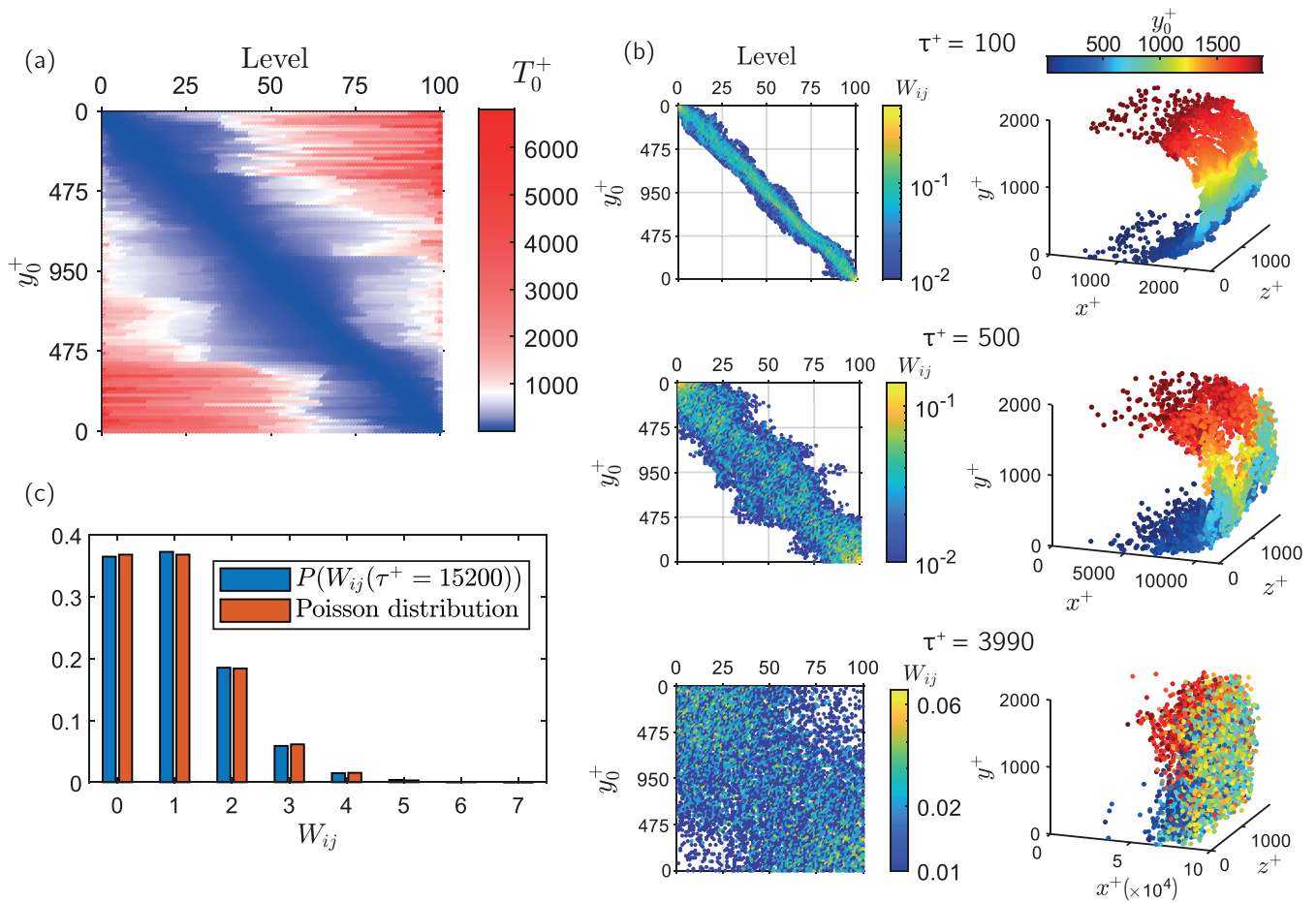}
\caption{Evolution of the transport matrix. Figure (a) shows the time $T_0^+$ at which each link is activated for the first time. Figures in (b) show the evolution of the network weight matrix and the corresponding distribution of tracers in the channel with their starting location color-coded; in the weight matrices, the start level is labeled with its distance from the nearest wall, while the end level is indicated by the node number. Figure (c) shows the weight distribution at $\tau^+ = \num{15200}$ and the Poisson distribution corresponding to a mean value of 1. \label{fig2}}
\end{figure*}

At the beginning of the mixing process only levels close in space are connected since tracers are mainly affected by streamwise advection\cite{iacobello2019lagrangian}. As time increases, tracers drift away from their starting position driven by the diffusive action of turbulence and new, longer range links are formed. In the end, tracers lose memory of their starting level and become thoroughly mixed across the channel.
The evolution of the network weight matrix $\mathbf{W}(\tau^+)$ provides a complete description of the vertical mixing phenomena occurring at any given time, as shown in figure \ref{fig2}.
Figure \ref{fig2}a shows the time $T_0^+$ at which each link is activated for the first time; the value of $T_0^+$ for each link $\lbrace i,\,j\rbrace$ is depicted in the corresponding matrix entry $i,\,j$. The starting level is labeled with its $y^+$ coordinate, while the ending one is labeled with the level number; the two ways of labeling are interchangeable. Also, since the distance of the level from the closest wall was reported instead of the $y^+$ coordinate, the $y^+$ values range in the interval $[0,\,\delta^+]$. As can be seen, the activation time depends both on the levels mutual distance (farther levels are connected at a later time) and, more notably, on the spatial position of the involved nodes. Indeed, links starting from a node located near one of the two walls tend to form later in time, as indicated by the higher values in the top right and bottom left portions of figure \ref{fig2}a. This highlights the fact that tracers released in those regions diffuse towards other parts of the channel more slowly than tracers from other levels to levels with a similar mutual distance. Besides, the opposite is not true, \textit{i.e.} particles starting in any other zone of the channel do not experience any significant delay in reaching the near-wall region.

Figure \ref{fig2}b shows the network weight matrix and the corresponding spatial distribution of particles for different times. For $\tau^+ = 100$ (figure \ref{fig2}b, first row), no significant mixing has taken place yet, so that only short range links are active. Tracers are distributed in a bow-like shape determined by the mean velocity profile; by this time, particles moving with the bulk velocity $V_b$ have traveled a downstream distance $\Delta x^+ =V_b\tau^+= 1985 \approx 2\delta$. At $\tau^+ = 500$ (figure \ref{fig2}b, second row), mixing is becoming effective, especially near the centerline of the channel, which in turns leads to the activation of new links; the mean downstream distance of particles is now $\Delta x^+ = 9930\approx 10.5\delta$. By $\tau^+ = 3990$, tracers appear dispersed in a cloud-like shape determined by diffusion (figure \ref{fig2}b, third row); not all links have been activated for the first time (as can be noted in figure \ref{fig2}a, all connections will be established for the first time after $\tau^+ \approx 6000$). The network weight matrix at this time still has larger values near the main diagonal, representing an ongoing influence of the mutual distance between levels in link generation. Tracers have traveled so far an average downstream distance $\Delta x^+ = 7.9\cdot 10^4 \approx 83.4\delta$. At the end of the simulation ($\tau^+ = \num{15200}$) all particles are completely mixed and connections are randomly distributed between levels. Tracers moving with the bulk velocity have traveled a downstream distance $\Delta x^+ = 3.02\cdot10^5 \approx 318\delta$. 

For very large times, when no memory of the initial tracer distribution is retained and mixing is complete, tracers are uniformly distributed, independently of the starting level. Neglecting the constant normalization factor $N_i$ in equation \eqref{eq:W} enables us to consider integer weights, which have unitary mean given by the ratio between the number of tracers ($N_p = \num{10000}$) and the number of possible links including self-loops ($N_l^2 = \num{10000}$). Since these non-normalized weights are independent and randomly distributed, they follow a Poisson distribution; the related probability distribution is
\begin{equation}
\label{eq:poi_pdf}
f(W_{ij}; \lambda) = \frac{\lambda^{W_{ij}} \mathrm{e}^{-\lambda}}{W_{ij}!},
\end{equation}
where $\lambda = 1$ is the mean value. The weight distribution was tested by the Kolmogorov-Smirnov test which confirmed that the data is Poisson distributed at the end of the simulation (with a significance level $\alpha = 0.05$). The empirical weight distribution at $\tau^+ = \num{15200}$ and the corresponding Poisson distribution for $\lambda = 1$ are shown in figure \ref{fig2}(c).

\begin{figure*}
\includegraphics[width = \textwidth]{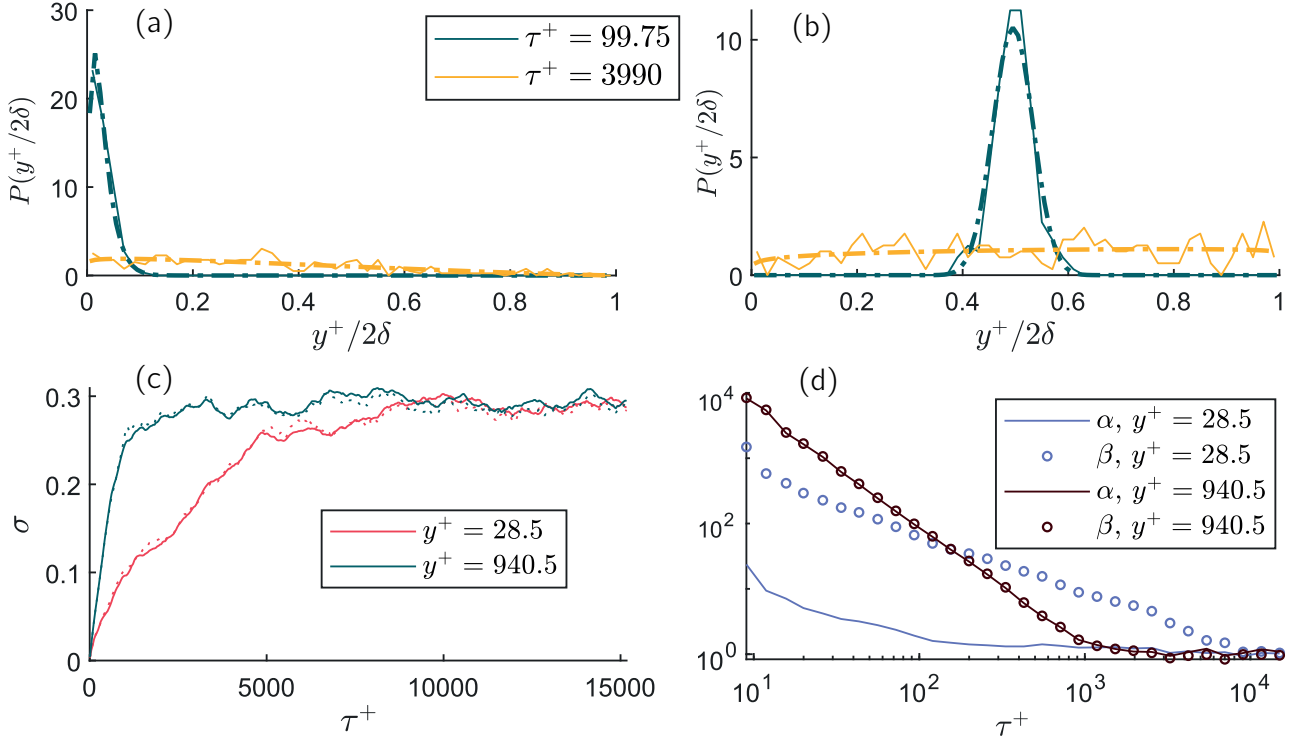}%
\caption{\label{fig3} Probability density function of tracer distribution after different time lags (solid lines), together with the fitted Beta distribution (dashed lines): (a) tracers starting from $y^+_0 = 28.5$ and (b) tracers starting from $y^+_0 = 940.5$. Both distributions are plotted as a function of the normalized channel height $y^+/2\delta$. (c) Standard deviation of the tracer distribution, both as measured (solid lines) and as obtained from the fitted Beta distribution (dotted lines); (d) $\alpha$ and $\beta$ parameters of the Beta distribution for two different levels.}
\end{figure*}

As stated in section \ref{sec:def}, each row $i$ of the network weight matrix $\mathbf{W}(\tau^+)$ describes the probability that a tracer starting from level $i$ is located in any other region of the channel after a time $\tau^+$. The shape of this distribution is heavily influenced by the release coordinate, $y_0^+$, of tracers, as can be seen in figures \ref{fig3}a and \ref{fig3}b for two exemplary cases corresponding to $y_0^+ = 28.5$ and $y_0^+ = 940.5$. The particle position distribution is significantly skewed by the presence of the wall for $y^+_0 = 28.5$, while it is almost symmetric for $y^+_0 = 940.5$; in the limit of very large times, all distributions become uniform on the $[0,\,2\delta]$ interval. The time evolution of the shape of this distribution provides information about the celerity at which tracers reach different regions. In particular, the standard deviation of the particle position measures the dispersion of particles; as can be seen in figure \ref{fig3}c, the standard deviation for a centerline level reaches its asymptotic value earlier in time than that of a near-wall level. After normalizing the $y^+$ coordinate with the channel height $2\delta^+$, we found that the asymptotic value for all standard deviations was nearly equal to $\sqrt{1/12}$; this is the expected value for an uniform distribution on the $[0,\,1]$ interval. The mean value of the distribution for each level (not shown here) trivially starts at the $y^+_0/(2\delta)$ normalized coordinate of that level and migrates to the centerline, \textit{i.e.} $1/2$, although the probability distributions related to levels near the walls reach the asymptote later in time. 

In order to further characterize the evolution of the previously defined distributions and to describe their evolution with a reduced set of parameters, we chose to fit these distributions with a Beta distribution, which has already been proposed as a model for diffusion problems \cite{chatwin1995turbulent}. The Beta distribution has a probability density function over the normalized channel height defined as
\begin{equation}
\label{eq:beta}
f\left(\frac{y^+}{2\delta}; \alpha,\beta\right) = \frac{\left(\frac{y^+}{2\delta}\right)^{\alpha-1} \left(1 - \frac{y^+}{2\delta}\right)^{\beta-1}}{\mathrm{B}(\alpha,\beta)} ,
\end{equation} 
where $\alpha$ and $\beta$ are two shape parameters and $\mathrm{B}(\alpha,\beta)$ is the beta function. This distribution accounts both for symmetrical ($\alpha = \beta$) and skewed ($\alpha\neq\beta$) distributions, as is the case for diffusion in the channel. Values of $\alpha$ and $\beta$, computed with a maximum likelihood estimate for the already considered levels are reported in figure \ref{fig3}d; the fitted Beta distribution is shown superimposed on the empirical distributions in figures \ref{fig3}a and \ref{fig3}b. We also verified the correspondence between the empirical distribution and the fitted one with a Kolmogorov-Smirnov test, which confirmed that the network weights follow a Beta distribution for $\tau^+\gtrsim 250$ (with a significance level $\alpha = 0.05$). 

As expected, $\alpha \approx \beta$ for $y^+_0 = 940.5$, while $\alpha \neq \beta$ for $y^+_0 = 28.5$; moreover, the $\alpha$ and $\beta$ parameters appear to follow a power law behavior with respect to time. Values for the first statistical moments can be calculated from $\alpha$ and $\beta$; figure \ref{fig3}c shows the measured standard deviation of particle position (solid lines) alongside the theoretical one (dotted lines), which is
\begin{equation}
\label{eq:betadev}
\sigma_B = \sqrt{\frac{\alpha\beta}{(\alpha+\beta)^2(\alpha+\beta+1)}},
\end{equation} 
indicating good agreement between the two empirical and fitted distributions.

\subsection{Network metrics analysis}
\label{sec:metric}
\subsubsection{Strength and degree}

\begin{figure*}
\includegraphics[width = \textwidth]{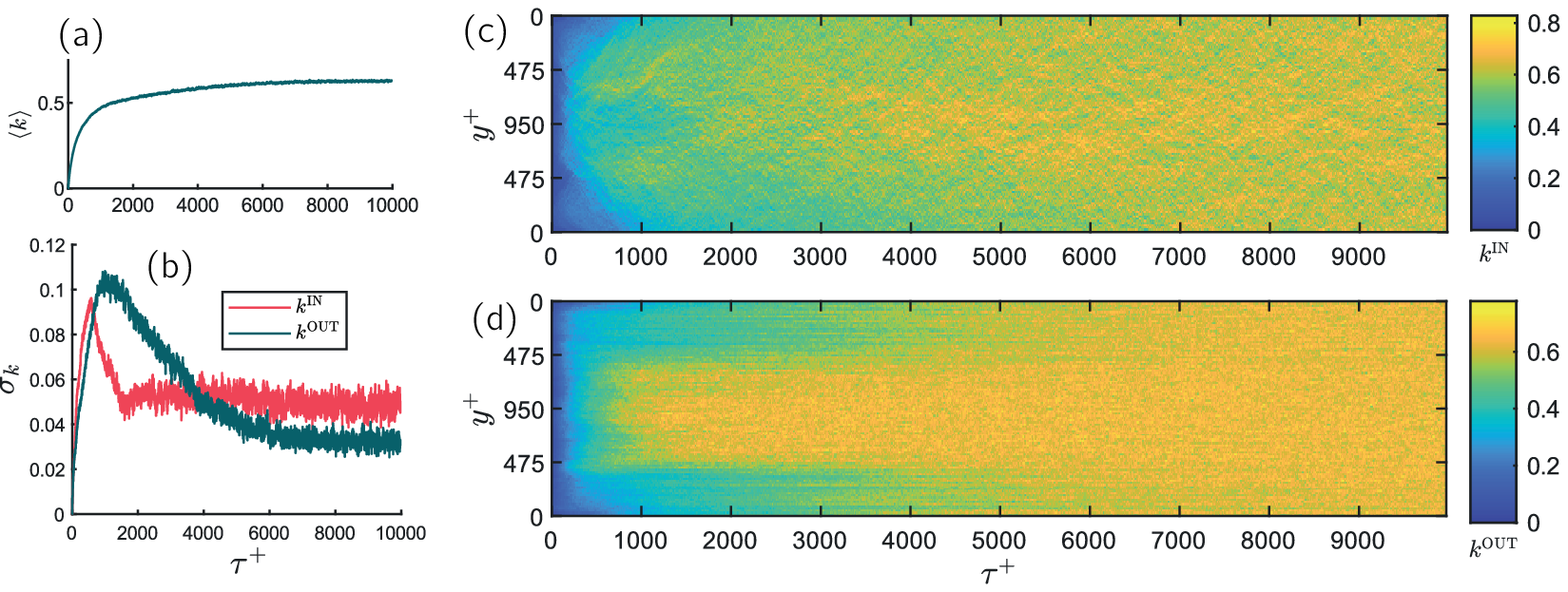}%
\caption{(a) Mean degree $\langle k \rangle$; (b) standard deviation of the ingoing and outgoing degree. (c) Ingoing degree, $k^{\text{IN}}$, and (d) outoing degree, $k^{\text{OUT}}$, shown for all levels and $\tau^+\in[0,\,\num{10000}]$. \label{fig4}}
\end{figure*}

This section is devoted to the analysis of two centrality measures, the strength and the degree, defined in section \ref{sec:net}. In particular, we will briefly describe the behavior of the strength of nodes, which is closely related to the concentration of tracers in each level. We will analyze in greater detail the degree centrality, which measures the activation of links neglecting their weight; we found that the degree is able to provide a simplified description of the spatial extent of mixing. We seek to highlight differences between nodes located at different $y^+$ coordinates and to evaluate the spatial inhomogeneities in the mixing process through the network formalism. 

The strength of nodes, being the sum of incident (ingoing or outgoing) link weights, is closely related to the values of the transport matrix $W_{ij}$. In more detail, the ingoing strength $s^{\uin}_i(\tau^+)$ is the sum of weights of links entering node $i$ at time $\tau$, which corresponds to the number of tracers present inside level $i$ at that time. Evaluating $s^{\uin}_i(\tau^+)$ enables us to monitor the concentration of tracers across the channel. On the other hand, the outgoing strength $s^{\uout}_i(\tau^+)$ is the summation of the rows of $W_{ij}$; since self-loops are included in the transport matrix, $s^{\uout}_i(\tau^+)$ is always equal to one. If we neglect self-loops and thus ignore tracers that do not change levels, the (modified) outgoing degree measures the number of tracers that are not present in their start level at time $\tau^+$; since tracers tend to leave their original level and disperse across the channel, $s^{\uout}_i(\tau^+)$ reaches a nearly unitary asymptote for all nodes in a very short time. 

The degree centrality of the transport network is, instead, the number of levels with which a node has established a connection by exchanging tracers. Levels with a higher degree are more active in the overall mixing process: high $k^{\uin}_i(\tau^+)$ values indicate that level $i$ has received tracers from a large number of levels, while a high outgoing degree $k^{\uout}_i(\tau^+)$ shows how tracers starting from level $i$ have entered a lot of levels. In the following, the degree of each node has been normalized with its maximum attainable value, which is 100, corresponding to a level which is exchanging particles with all other levels including itself.

Figure \ref{fig4}a shows the mean degree $\langle k \rangle$ for $\tau^+ \in [0,\,\num{10000}]$, a time after which no variations are observed (the mean $\langle \cdot\rangle$ is calculated over the degree of all levels); it should be noted that $\langle k^{\uin}\rangle = \langle k^{\uout}\rangle$ by the definition of degree in equation \eqref{eq:k}. The asymptotic value of $\langle k \rangle$ is approximately $\num{0.63}$; recalling that the non-normalized network weights are Poisson distributed for large times, we can calculate the probability that a weight is zero by substituting in equation \eqref{eq:poi_pdf} $W_{ij} = 0$ and $\lambda  = 1$, obtaining $P(W_{ij} = 0) = f(0; 1) = 1/\mathrm{e}\approx \num{0.37}$; since if a weight is zero there is no link between two nodes, the expected mean degree at a large time $\tau^+$ is equal to the fraction of non-zero weights in the network, namely $\langle k \rangle = 1 - 1/\mathrm{e} \approx {0.63}$, which is also the measured value. While this value is strictly dependent on the ratio between the number of tracers and the number of possible links, the overall behavior of the degree centrality is invariant to changes in the number of particles.

The standard deviation of the ingoing and outgoing degree is shown in figure \ref{fig4}b. The standard deviation of the ingoing degree reaches a maximum at about $\tau^+ = 500$, while the standard deviation of $k^{\uout}$ reaches its maximum shortly after; both degrees subsequently decrease and reach an asymptote. The standard deviation of the outgoing degree reaches its asymptotic value at $\tau^+\approx6000$, while the ingoing degree reaches its asymptote (which has a larger value) in a shorter time ($\tau^+\approx 2000$). The sharp increase in standard deviation experienced during the transient phase indicates an inhomogeneous increase of the degree across the channel and, therefore, the presence of levels that form a smaller/higher number of connections than others. 

Again, the asymptotic values of the standard deviation of the degree can be obtained using the Poisson weight model by generating a random transport matrix with Poisson distributed weights and the sum of each row fixed and equal to $N_i = 100$. We obtained a standard deviation for the ingoing degree equal to 0.044 (the measured value was 0.047); for the outgoing degree, the standard deviation was 0.032 for the random model and 0.031 for the measured one.

To further characterize the spatial inhomogeneities during the early stage of particle dispersion, we show the ingoing and outgoing degree through time and decomposed by levels, in figures \ref{fig4}c and \ref{fig4}d. During the very early phase the ingoing degree rises more intensely near the centerline, indicating that these levels tend to form a large number of ingoing connections and are thus receiving tracers from a variety of different and, possibly, distant levels. In particular, the ingoing degree grows faster than its surroundings at a distance from the walls $y^+\approx 475$; the spikes in the degree $k^{\uin}(\tau^+)$ appear at a time for which its standard deviation is maximal. Also, the outgoing degree presents a sharp difference between the two near-wall zones and the core region, which are again demarcated at $y^+\approx 475$. This behavior hints at the presence of a neat separation of mixing properties inside the channel; near-wall regions form a somewhat secluded zone, which transfers a reduced amount of tracers to the external flow. These inhomogeneities are more evident for $\tau^+$ values between 1000 and 2000, which corresponds to the time at which the maximum of the standard deviation of the outgoing degree occurs. 

It is also notable that tracers do not experience the same difficulty while entering the near wall regions, as the ingoing degree shows; this hints that tracers released at a low $y^+_0$ coordinate become trapped near their starting location for a long time. The presence of a stable, long-lasting inhomogeneity of the degree is the reason why the asymptote of the standard deviation of $k^{\uout}$ is reached later in time. We provide further evidence of this trapping of tracers with additional analyses based on the network features.

\subsubsection{Link length and link duration}
\begin{figure*}
\includegraphics[]{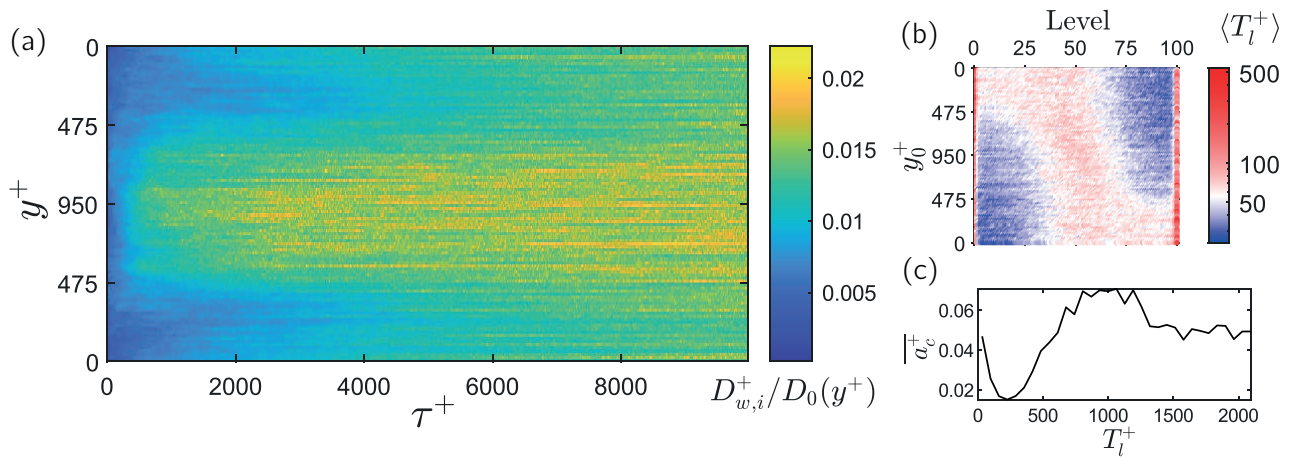}
\caption{(a) Weighted link length $D_{w,i}^{+,\uout}/D_0$, averaged for each node and corresponding to outgoing links; (b) mean temporal duration of each link $\langle T_l^+ \rangle$ (the highest values are found for end levels 1 and 100); (c) Trendline of the mean centripetal acceleration versus link duration\label{fig5}}
\end{figure*}

Since the time-dependent transport network is embedded in physical space, useful information can be extracted by evaluating the mean weighted link length $D_{w,i}^+$ introduced in equation \eqref{eq:wedgel}, which in our case represents the spatial distance between interacting levels weighted by the number of exchanged tracers, averaged for each node. The link length is a physical measure of the magnitude of the displacement of particles from their starting location; unlike topological measures, such as the characteristic network length, it is not related to paths. We use this metric for two main reasons: first, to obtain a condensed measure of both the mutual spatial distance of interacting levels and of the weight of the interaction. Secondly, we hypothesize that, while the majority of tracer exchanges happens locally, some infrequent events are present, in which a large number of particles moves between spatially distant levels. The use of $D_{w,i}^+$ is suitable to highlight this kind of extreme behavior.

Figure \ref{fig5}a shows the outgoing and weighted link length $D_{w,i}^{+,\uout}$ of all the levels inside the channel for a wide range of times. We normalized the value of the link length of each level with a factor $D_0$ dependent on the $y^+$ value of that level and equal to
\begin{equation}
\label{eq:normfac}
D_0(y^+) = \frac{1}{2\delta}\int_0^{2\delta} \lvert{s - y^+}\rvert\mathrm{d}s = \delta - y^+ + \frac{(y^{+})^2}{2\delta}.
\end{equation}
We do so to divide the value of $D_{w,i}^{+,\uout}$ by the average physical distance between level $i$ and the rest of the channel; this enables us to account for the fact that near-wall levels have higher $D_{w,i}^{+,\uout}$ due to their position (since they are able to form longer distance links). Figure \ref{fig5}a shows that $D_{w,i}^{+,\uout}/D_0$ rises faster in the core region and exhibits a similar distinction between the near-wall and the external region as found in the outgoing degree (figure \ref{fig4}c). Again, the central region is found to be more active in the mixing process; only for very large times $\tau^+$ does the weighted link length of near-wall levels rise significantly. As time grows, it can be noted that some nodes have a larger $D_{w,i}^{+,\uout}/D_0$ than their closest neighbors and that this behavior is quite persistent in time; this determines the presence of streaks in figure \ref{fig5}a, especially for large times. Some of these streaks appear to last for times of the order of $1000\nu/u^2_{\tau}$. We hypothesize that this behavior is related to the occurrence of connections between distant levels which have a long duration, and are caused by tracers that spend a long time trapped away from their starting level.

It is reasonable to assume that persistent and spatially confined turbulent structures could cause tracers to remain trapped in small regions, generating links that have such long durations. Each link has a temporal duration $T_l^+$, equal to the total time for which it is active at consecutive time instants; this corresponds to the time during which a tracer stays in the same level without leaving it. Figure \ref{fig5}b shows the mean duration of links $\langle T^+_l\rangle$ for each pair ${i,\,j}$ of nodes, calculated through the entire evolution of the trajectories. The mean $\langle\cdot\rangle$ here is calculated over the duration of links between the same pair of nodes. Values of $\langle T^+_l\rangle$ vary greatly with respect to the end level, while they are almost constant for different start levels; higher values of $\langle T^+_l\rangle$ are found for end levels near the centerline and, especially, for the two levels closest to the walls, \textit{i.e.} levels 1 and 100. The mean link duration is more than an order of magnitude larger for links ending in the wall-adjacent levels than for other connections.  
This indicates that tracers which enter the levels nearest to the wall are likely to be trapped and stay in that levels for a longer amount of time than those which enter other levels; this happens independently of their starting level. 

The low $y^+$ regions in a channel are usually home to a number of coherent structures, such as hairpin and horseshoe vortices. These structures are known to induce helical motion and therefore trap tracers in small-radius vortices \cite{polanco2019lagrangian}, as is exemplified by the particle motion analyzed here.
A typical measure for helical motion in turbulent flows is the centripetal acceleration experienced by particles \cite{toschi2005acceleration}. We calculated the acceleration $\mathbf{a}$ of tracers using a second-order finite difference and computed the modulus of the acceleration centripetal component as
\begin{equation}
\label{eq:centracc}
a_c = \lvert\mathbf{a}\times \hat{\mathbf{v}}\rvert,
\end{equation}
where $\hat{\mathbf{v}} = \mathbf{v}/\lvert\mathbf{v}\rvert$ is unit vector pointing in the direction of the velocity.
While a link $W_{ij}(\tau)$ is active, the tracers generating that link experience a centripetal acceleration which varies with time. We averaged the centripetal acceleration of particles for the entire duration of a link, obtaining the mean centripetal acceleration $\overline{a^+_c}$ of each link; we also calculated the duration $T^+_l$ of these links. To highlight the correlation between the centripetal acceleration and the duration of links, we show in figure \ref{fig5}c a trendline of $\overline{a^+_c}$ versus $T^+_l$, computed binning the $T^+_l$ values in equal intervals and averaging the corresponding $\overline{a^+_c}$ values.

Three distinct behaviors can be identified: for relatively brief link durations ($T^+_l \approx 250$) the mean centripetal acceleration is low, indicating that spiraling motion is not particularly important in this kind of connections. On the other hand, tracers participating in both very brief ($T^+_l\lesssim 150$) and very long-lasting ($T^+_l\gtrsim 500$) links have significantly higher mean centripetal acceleration. In very brief links, tracers may possibly experience short bursts of acceleration which move them outside their current level and quickly end the connection. On the other hand, very long-lasting links are generated by tracers which stay in the same, confined, region in space and go through strong spiraling motion.
The majority of links (about 99\% of the total number of links) is very brief; very long-lasting links, instead, are extreme events which occur far less often (736 out of approximately $5\cdot10^6$ links) than all other connections. The transport network highlights the relationship between the intense helical motion of tracers retained in spatially confined structures, the temporal persistence of these structures (figure \ref{fig5}c) and their vicinity to the walls (figure \ref{fig5}b). 

\subsection{Communities}
\label{sec:com}

\begin{figure}
\includegraphics[width = 8.6cm]{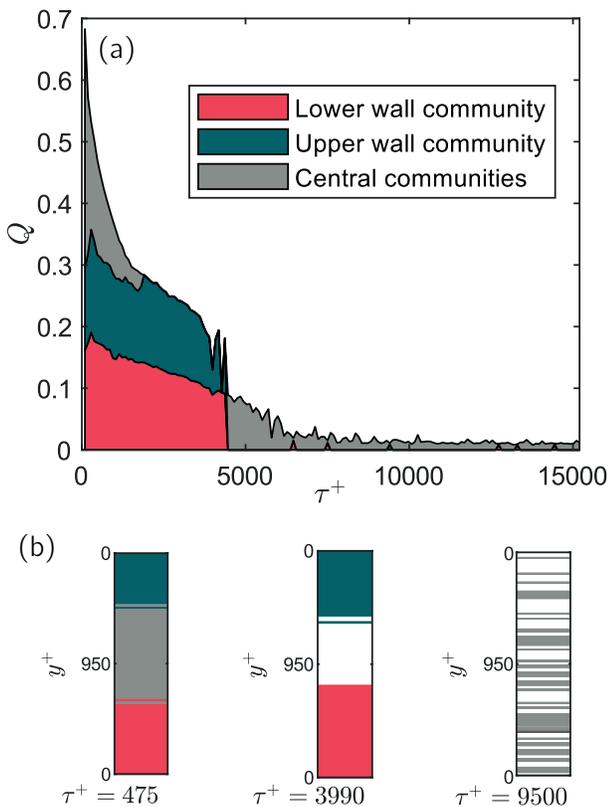}
\caption{(a) The modularity of the network is shown for different times and decomposed into the contributions given by the near-wall (top and bottom) and all the central communities together; (b) Partitioning for three different times is shown: for very short times, two near-wall regions are present, while in the channel core there is a large number of small communities, which we all indistinctly show in gray; for intermediate times, only the near-wall communities remain (the core region presents no community structure and is shown in white); in the end, all communities break up and no modular structure can be identified (only small communities are present, shown in gray; still, their contribution to the modularity is negligible). \label{fig6}}
\end{figure}

The degree centrality and the weighted link length indicate the presence of separate regions inside the channel, with different mixing properties. 
In order to better identify this inhomogeneity, we perform a community detection in the network. Unlike other approaches based on Lagrangian particle tracking \cite{schlueter2017coherent}, we do not search for sets of tracers which move coherently across the channel, but rather partition the turbulent channel into regions which exchange a reduced number of tracers between themselves.
To do so, we use the algorithm introduced by \textcite{girvan2002community}, which we will briefly describe here. The main idea is to calculate the betweenness of all network links; those with a higher betweenness are bottlenecks in communication between different regions of the network. In our case, high betweenness links play an important role in transferring tracers between separate zones of the channel. Since our network has integer weights up to a constant and the weights indicate the importance of a connection (in opposition to its cost, as is the case with many other networks), we can employ the extension introduced by \textcite{newman2004weighted}. 
Starting from a single-time transport network $W(\tau^+)$, we proceed as follows: we calculate the betweenness $B_{ij}$ of each link and divide it by its weight; then, we remove the link with the highest $B_{ij}/W_{ij}$ value from the network. To analytically partition the network, we employ Tarjan's algorithm \cite{tarjan1972depth} to find strongly connected components - \textit{i.e.} components of the graph in which there is a bidirectional path between all node pairs - and we calculate the modularity $Q$ of the complete network when partitioned with these components. This step is repeated until no link remains in the network; the partition which maximizes the modularity is chosen as the most suitable division of the network into communities \cite{newman2004detecting}.

This algorithm, applied to networks built at different times $\tau^+$, identifies two long lasting communities located near each of the channel walls; also, some smaller communities are found near the centerline, which rapidly break up and disappear. Figure \ref{fig6}a shows the modularity of the network over time, decomposed into the contribution given by the two near-wall regions and all the other communities. As can be seen, the majority of the modularity contribution comes from the two near-wall communities until $\tau^+\approx 4000$, when these partitions dissolve and no modular structure can be identified anymore. At a similar time, the outgoing degree (figure \ref{fig4}d) starts to become more homogeneous across the channel height; it is the end of the transient phase of the mixing process.
The spatial distribution of the communities during different phases of the flow is shown in figure \ref{fig6}b. While the central communities are possibly related to the local nature of particle movement during the very first phase of the mixing process (as in figure \ref{fig2}b, first and second rows), the top and bottom communities are coherent in time and space and indicate the presence of two secluded regions, that is regions at $y^+$ coordinates lower than about $475 = \delta^+/2$, which exchange a reduced number of particles with the centerline zone. The communities located in the core region are initially small (we show all of them indistinctly in gray in figure \ref{fig6}b) and quickly dissolve. 

In the end, the mixing process leads to the dissolution of all communities; yet, shortly after their release tracers located near the walls are less likely to move towards the rest of the channel. Finally, we compared this method to other partitioning methods employed in analogous cases \cite{sergiac}, finding that it yields similar results.

\section{Conclusions}
\label{sec:concl}

We analyzed turbulent mixing in a wall-bounded flow from the Lagrangian perspective of advected fluid tracers using a network-based approach. While conceptually simple and easily implementable, treating the dispersion of particles as a set of interactions between discrete zones of a fluid domain highlights the richness of detail of the mixing process. Simple network metrics, such as the degree centrality, are able to measure the extent of mixing at different times and in different locations of a channel flow. In particular, during the transient phase, not all tracers mix together evenly; those released near one of the two channel walls move more slowly towards the center of the channel.

We introduced a metric, the mean weighted link length $D_w$, which is capable of highlighting extreme events experienced by tracers trapped in small regions. The spatial and temporal features of network links are closely related to the characteristics of Lagrangian trajectories; we have shown that long duration links are mostly located very close to the channel walls and are generated by tracers experiencing strong helical motion. Links involving trapped particles last up to an order of magnitude longer than other connections, while the associated centripetal acceleration is up to three times stronger. 

The application of established partitioning techniques enabled us to further refine the intuition originating from the degree analysis, \textit{i.e.} during the first phase of the flow evolution mixing does not involve the channel in its entirety. The partitioning offered by the betweenness approach compares well with the other transport network features and offers an immediate quantification of the spatial extent of mixing. Still, the features of this approach should be further assessed, especially by a comparison with other measures of mixing.

Using the network approach, we were able to assess the main features of the mixing process: we determined the existence and temporal duration of a transient phase in which the initial location of the tracers is still relevant and the role of the channel walls in forcing the dispersion of particle locations to confined regions of the domain. As noted in section \ref{sec:weight}, the variance of the distribution of particles released near the walls reaches its asymptote (corresponding to a uniform distribution of tracers) later in time than that of particles starting away from the solid boundaries. In section \ref{sec:metric} we have shown that during the initial transient the outgoing degree, which corresponds to the number of different locations reached by the particles, is about 30\% lower in the near wall region. Moreover, the demarcation between the near wall zones and the channel core is sharp and located consistently at $y^+ = \delta/2$ until the end of the transient period, which extends until $\tau^+\approx 4000$. The division of the network into communities showed the presence of two main clusters of nodes, located adjacent to each wall, which give the main contribution to the modularity $Q$ of the network. The modularity reduces significantly after $\tau^+  =4000$, as the transient ends and mixing becomes effective in transferring scalars across the entire channel.

While we did not investigate directly the effects of the Reynolds number, by choosing to perform our analysis at $Re_{\tau} = 950$ we achieved a sufficient separation of scales and a sufficiently quick mixing to hypothesize a similar behavior of the transport network at higher Reynolds numbers.
The transport network was applied to a simple geometry, but it can also be used to describe different and more complex flow cases, if an adequate domain partition is chosen in order to define network nodes. Also, since we were mainly interested in vertical mixing, we restricted the analysis to a one-dimensional node layout, but this could be changed to investigate the motion of particles in more than one direction. This could be of use in investigating the efficiency of mixing in complex geometries or the tendency of inertial particles to cluster.

The network representation of fluid trajectories provides a bridge between the analysis of the behavior of single particle motion and the study of the overall mixing process. This is a result of the capabilities of complex networks, which are able to sum up the separate behaviors of a vast number of different agents (in our case, the particles moving in a turbulent flow) and bring out the emerging collective features in great detail (for us, the mixing phenomenon as a whole).

\appendix*
\section{Direct numerical simulation}
\label{sec:dns}

The Navier-Stokes equations for incompressible flow are solved in their rotation form
\begin{equation}
\label{ns1}
\nabla\cdot \mathbf{v} = 0,
\end{equation}
\begin{equation}
\label{ns2}
\frac{\partial \mathbf{v}}{\partial t} + \frac{1}{\rho}\nabla P = \mathbf{f} - \boldsymbol\omega\times \mathbf{v} + \nu \nabla^2\mathbf{v},
\end{equation}
where $\rho$, $\mathbf{v}$ and $\nu$ are the mass density, the velocity vector and the kinematic viscosity, $P$ is the total pressure and $\boldsymbol\omega$ is the vorticity. The equations \eqref{ns1} and \eqref{ns2} are made non-dimensional using the mass density, the friction velocity $u_{\tau}$ and the half channel height $\delta$. The frictional Reynolds number $Re_{\tau} = 950$ is kept fixed by prescribing the mean driving force per unit mass $\mathbf{f}$. The bulk velocity is not fixed and its mean value is $V_b^+ = V_b/u_{\tau} = 19.8$, therefore the bulk Reynolds number is about $\num{18900}$.
Equations \eqref{ns1} and \eqref{ns2} were solved in a rectangular box of size $2\pi\delta\times 2\delta \times \pi\delta$, with periodic boundary conditions in the two homogeneous directions ($x$ and $z$); on the walls the no-slip condition $\mathbf{v} = 0$ was imposed. A pseudo-spectral method was used, with a Fourier-Galerkin approach in the homogeneous directions and a Chebyshev-tau method in the wall-normal. Non linear terms were calculated in physical space through Fast Fourier Transform and with the application of the $3/2$ rule; they were explicitly integrated in time with the second order Runge-Kutta method, while linear terms were advanced in time through an implicit Crank-Nicolson scheme. The time step used in the simulation is $\Delta t^+_{\text{DNS}} = 0.095$.

To integrate the Lagrangian trajectories of the seeded tracers the same explicit second order Runge-Kutta method as employed in the DNS was used. The velocity $\mathbf{v}(\mathbf{x}(\mathbf{x}_0, t),t)$ at the tracer location was calculated with trilinear interpolation using the velocity field after the Fourier transform to physical space, so that the number of points in the two periodic directions is increased by a factor of $3/2$. Although higher order schemes could have been employed to achieve better precision in the determination of single trajectories, the accuracy of statistical particle results is unaffected \cite{choi2004lagrangian,van2013optimal,kuerten13}.
The trajectories of seeded tracers are followed through the total simulation time $T^+ = \num{15200}$ and their position is recorded with a time step $\Delta t^+ = 0.475$, resulting in $N_t = \num{32000}$ snapshots of the particle positions.

\begin{acknowledgments}
This work was sponsored by NWO Exacte en Natuurwetenschappen (Physical
Sciences) for the use of supercomputer facilities, with financial support from the
Netherlands Organization for Scientific Research, NWO.
\end{acknowledgments}

\onecolumngrid

\end{document}